\documentclass[english]{article}
\usepackage[T1]{fontenc}
\usepackage[latin9]{inputenc}
\usepackage{setspace}
\usepackage{amssymb}
\usepackage{esint}

\makeatletter
\newcommand{\lyxaddress}[1]{
\par {\raggedright #1
\vspace{1.4em}
\noindent\par}
}

\makeatother

\usepackage{babel}

\begin{document}
\begin{doublespace}

\title{Holographic principle and large scale structure in the universe}
\end{doublespace}

\begin{doublespace}

\author{T. R. Mongan}
\end{doublespace}

\maketitle
\begin{doublespace}

\lyxaddress{84 Marin Avenue, Sausalito, California 94965 USA; tmongan@gmail.com}
\end{doublespace}
\begin{abstract}
\begin{singlespace}
A reasonable representation of large scale structure, in a closed
universe so large it's nearly flat, can be developed by extending
the holographic principle and assuming the bits of information describing
the distribution of matter density in the universe remain in thermal
equilibrium with the cosmic microwave background radiation. The analysis
identifies three levels of self-similar large scale structure, corresponding
to superclusters, galaxies, and star clusters, between today's observable
universe and stellar systems. The self-similarity arises because,
according to the virial theorem, the average gravitational potential
energy per unit volume in each structural level is the same and depends
only on the gravitational constant. The analysis indicates stellar
systems first formed at $z\approx62$, consistent with the findings
of Naoz \emph{et al}, and self-similar large scale structures began
to appear at redshift $z\approx4$. It outlines general features of
development of self-similar large scale structures at redshift $z<4$.
The analysis is consistent with observations for angular momentum
of large scale structures as a function of mass, and average speed
of substructures within large scale structures. The analysis also
indicates relaxation times for star clusters are generally less than
the age of the universe and relaxation times for more massive structures
are greater than the age of the universe. \end{singlespace}

\end{abstract}

\begin{doublespace}

\section{Introduction}
\end{doublespace}

\begin{doublespace}
Formation of large scale structure in the universe is an important
problem in cosmology \cite{key-1}, and the heuristic Press-Schechter
excursion set model has been considered the only viable analytic approach
to formation of large scale structure \cite{key-2}. In contrast,
this analysis extends the holographic principle \cite{key-3} to consider
formation of large scale structures, and stellar systems comprising
those structures, in a closed Friedmann universe so large it's nearly
flat. That may be a reasonable approximation to our universe.

In this analysis, $\rho_{r}(z)$ is the cosmic microwave background
(CMB) radiation density at redshift $z$, where $\rho_{r}(z)=(1+z)^{4}\rho_{r}(0)$
and the mass equivalent of today's radiation energy density $\rho_{r}(0)=4.4\times10^{-34}$g/cm$^{3}$
\cite{key-4}. Correspondingly, $\rho_{i}(z)$ is the matter density
within structural level $i$ at redshift $z$ and $\rho_{0}(0)$ is
today's matter density in the universe as a whole. If the Hubble constant
$H_{0}=71$ km/sec Mpc, the critical density $\rho_{crit}=\frac{3H_{0}^{2}}{8\pi G}=9.5\times10^{-30}$g/cm$^{3}$
where $G=6.67\times10^{-8}$ cm$^{3}$g$^{-1}$sec$^{-2}$, and $c=3.00\times10^{10}$cm
sec$^{-1}$. Assuming the universe is dominated by vacuum energy resulting
from a cosmological constant $\Lambda$, matter accounts for about
26\% \cite{key-5} of the energy in today's universe. So, $\rho_{0}(0)=0.26\rho_{crit}=2.5\times10^{-30}$g/cm$^{3}$
and the vacuum energy density $\rho_{v}=\left(1-0.26\right)\rho_{crit}=7.0\times10^{-30}$g/cm$^{3}$.
The cosmological constant $\Lambda=\frac{8\pi G\rho_{v}}{c^{2}}=1.3\times10^{-56}$cm$^{2}$
and there is an event horizon in the universe at radius $R_{H}=\sqrt{\frac{3}{\Lambda}}=1.5\times10^{28}$cm.
Therefore, the mass $M_{u}$ of the observable universe is about $M_{u}=\frac{4}{3}\pi R_{H}^{3}\rho_{0}(0)=3.6\times10^{55}$g. 

According to the holographic principle \cite{key-3}, the number of
bits of information available on the light sheets of any surface with
area $a$ is $\frac{a}{4\delta^{2}ln\left(2\right)}$, where $\delta=\sqrt{\frac{\hbar G}{c^{3}}}$
is the Planck length and $\hbar$ is Planck's constant. So, only $N=\frac{\pi R_{H}^{2}}{\delta^{2}ln\left(2\right)}=4.0\times10^{122}$
bits of information on the event horizon will ever be available to
describe all physics within the event horizon in our universe, The
average mass per bit of information in the universe is $\left(3.6\times10^{55}g\right)/\left(4.0\times10^{122}\right)=9.0\times10^{-68}g$
and the holographic principle indicates the total mass of the universe
relates to the square of the event horizon radius by $M_{u}=fR_{H}^{2}$,
where $f=0.16$g/cm$^{2}$. 

In a closed universe, there is no source or sink for information outside
the universe, so the total amount of information in the universe remains
constant. Also, after the first few seconds of the life of the universe,
energy exchange between matter and radiation is negligible compared
to the total energy of matter and radiation separately \cite{key-6}.
So, in a closed universe, the total mass of the universe is conserved
and the average mass per bit of information is constant. This suggests
an extension of the holographic principle indicating the information
describing the physics of an isolated gravitationally-bound astronomical
system of total mass $M$ is encoded on a spherical holographic screen
with radius $R=\sqrt{\frac{M}{0.16}}$cm around the center of mass
of the system. 
\end{doublespace}

\begin{doublespace}

\section{Assumptions}
\end{doublespace}

\begin{doublespace}
In a closed universe, a hierarchical self-similar description of the
development of large scale structure in the universe can be obtained
based on four assumptions:
\end{doublespace}
\begin{enumerate}
\begin{doublespace}
\item Extend the holographic principle by assuming all information necessary
to describe an isolated astronomical structure of mass $M$ is available
on the light sheets of a holographic spherical screen with radius
$R=\sqrt{\frac{M}{0.16}}$cm around the center of mass of the structure,
so the average matter density within the spherical screen is $\rho_{M}=\frac{0.16R^{2}}{\frac{4}{3}\pi R^{3}}=\frac{0.12}{\pi R}$g/cm$^{3}$.
\item Assume the bits of information on the holographic spherical screens
surrounding isolated astronomical structures are in thermal equilibrium
with the CMB radiation.
\item Assume structures at any given self-similar structural level range
in mass from the Jeans' mass at that level down to the Jeans' mass
for the next finer level of structure.
\item Assume the number of structures of mass $m$ in any structural level
$i$ is $\frac{K}{m}$, where $K$ is constant, so the amount of information
in any mass bin (proportional to $\frac{K}{m}m$) is the same in all
mass bins. This is consistent with the $\frac{1}{m}$ behavior of
the mass spectrum in the Press-Schechter formalism.\end{doublespace}

\end{enumerate}
\begin{doublespace}

\section{Analyses}
\end{doublespace}

\begin{doublespace}
Based on these assumptions, the following analysis identifies three
levels of self-similar large scale structure (corresponding to superclusters,
galaxies, and star clusters) between today's observable universe and
stellar systems. Those self-similar large scale structures can be
seen as gravitationally-bound systems of $n$ widely separated units
of the next lower structural level in a sea of cosmic microwave background
photons. In this approach, today's speed of pressure waves affecting
matter density at structural level $i$ is $c_{si}(0)=\frac{2c}{3}\sqrt{\frac{\rho_{r}(0)}{\rho_{i}(0)}}$
\cite{key-7} , and the corresponding Jeans' length $L_{i+1}(0)=c_{si}(0)\sqrt{\frac{\pi}{G\rho_{i}(0)}}$
\cite{key-7}. In today's universe, $c_{s0}=2.7\times10^{8}$cm/sec,
and the first level (supercluster) Jeans' length $L_{1}(0)=1.2\times10^{27}$cm.
The first level Jeans' mass, the mass of matter within a radius one
quarter of the Jeans' wavelength $L_{1}(0)$, is $M_{1}(0)=\rho_{0}(0)\frac{4}{3}\pi\left(\frac{L_{1}(0)}{4}\right)^{3}=2.6\times10^{50}$g.
All scales smaller than the Jeans' wavelength are stable against gravitational
collapse, and the radius of the spherical holographic screen for the
first level Jeans' mass is $R_{1}=4.1\times10^{25}$cm. The matter
density within the spherical holographic screen for the first level
Jeans' mass is $\rho_{1}(0)=\frac{0.16R_{1}^{2}}{\frac{4}{3}\pi R_{1}^{3}}=\frac{0.12}{\pi R_{1}}=9.1\times10^{-28}$g/cm$^{3}$.
Then, $c_{s1}=1.4\times10^{7}$cm/sec within the first level Jeans'
mass, the second level (galaxy) Jeans' length is $L_{2}(0)=3.2\times10^{24}$cm,
and the second level Jeans' mass is $M_{2}(0)=\rho_{1}(0)\frac{4}{3}\pi\left(\frac{L_{2}(0)}{4}\right)^{3}=1.9\times10^{45}$g.
Continuing in this way, the third level (star cluster) Jeans' mass
$M_{3}(0)=1.4\times10^{40}$g, the fourth level (stellar system) Jeans'
mass $M_{4}(0)=1.0\times10^{35}$g, and $\frac{M_{1}(0)}{M_{u}}=\frac{M_{2}(0)}{M_{1}(0)}=\frac{M_{3}(0)}{M_{2}(0)}=\frac{M_{4}(0)}{M_{3}(0)}=7.3\times10^{-6}$.
The hierarchy of large scale structure stops with star clusters, because
stellar systems cannot be treated as systems consisting of $n$ widely
separated subelements in a sea of cosmic microwave background photons.

Identify superclusters as structures with masses between the first
and second level Jeans' masses, galaxies as structures with masses
between the second and third level Jeans' masses, and star clusters
as structures with mass between the third and fourth level Jeans'
masses. Then, the universe can be seen successively as an aggregate
of superclusters, an aggregate of galaxies, an aggregate of star clusters,
or an aggregate of stellar systems. The Jeans' masses identify each
structural level, but a mass distribution is needed to estimate the
number of entities in each structural level and the average mass of
structures at that level. Using the assumed $\frac{K}{m}$ behavior
of the mass spectrum, the number of superclusters in the universe
is $n=\intop_{7.3\times10^{-6}M_{1}}^{M_{1}}\left(\frac{K}{m}\right)dm=11.8K$
and the mass of the universe relates to the aggregate of supercluster
masses by $M_{u}=\intop_{7.3\times10^{-6}M_{1}}^{M_{1}}m\left(\frac{K}{m}\right)dm\approx KM_{1}$.
So, $K=\frac{M_{u}}{M_{1}}$, the average mass of a supercluster $\overline{M_{1}}=\frac{M_{u}}{n}=\frac{M_{1}}{11.8}=2.2\times10^{49}$g
and the mass of the universe is the number of superclusters times
the average supercluster mass. There are $n=\intop_{7.3\times10^{-6}M_{2}}^{M_{2}}\left(\frac{K}{m}\right)dm=11.8K$
galaxies in a first level Jeans' mass, and the first level Jeans'
mass is the aggregate of the galaxy masses within that Jeans' mass,
so $M_{1}=\intop_{7.3\times10^{-6}M_{2}}^{M_{2}}m\left(\frac{K}{m}\right)dm\approx KM_{2}$.
Then, $K=\frac{M_{1}}{M_{2}}$, and the average galaxy mass $\overline{M_{2}}=\frac{M_{1}}{n}=\frac{M_{2}}{11.8}=1.6\times10^{44}$g.
A similar analysis gives an average star cluster mass of $1.2\times10^{39}$g,
and these results are consistent with observations \cite{key-8},
\cite{key-9}, \cite{key-10}.

Down to the third (star cluster) structural level, the total number
$n=11.8K=1.6\times10^{6}$ of next lower level substructures inside
the holographic screens for the Jeans' length at each structural level
are the same as the total number of superclusters in the observable
universe. Furthermore, considering the large scale structures within
the universe, there are $1.4\times10^{5}$ average mass galaxies in
an average mass supercluster, $1.4\times10^{5}$ average mass star
clusters in an average mass galaxy and (if the average stellar system
mass is 4.3 times the solar mass) $1.4\times10^{5}$ average mass
stellar systems in an average mass star cluster. To understand the
self-similarity (scale invariance) of large scale structures within
the universe, consider gravitationally-bound systems of $n$ entities
with mass $m$ and total mass $M=nm$. For structures with $n\approx10^{5}$,
the substructure mass $m$ is much less than the mass $M$ of the
next highest level of structure. From the virial theorem, the gravitational
potential energy of the systems is $V_{G}=-\frac{GM^{2}}{2R}.$ The
extended holographic principle indicates the information needed to
describe gravitationally-bound astronomical systems of total mass
$M$ consisting of empty radiation-filled space and $n$ smaller entities
with mass $m\ll M$ is available on a spherical holographic screen
of radius $R=\sqrt{\frac{M}{0.16}}$ surrounding the system. Then,
the gravitational potential energy of the structure of mass $M$ within
the holographic screen is $V_{G}=-\frac{GM^{2}}{2R}=-\frac{G(0.16)^{2}R^{3}}{2}$,
so self-similarity (scale invariance) of large scale structures occurs
because the average gravitational potential energy per unit volume
at each structural level depends only on the gravitational constant
and is identical for all levels of large scale structure.

Now consider development of large scale structure at $z>0.$ Stellar
systems are the basic elements of self-similar large scale structures
(star clusters, galaxies, superclusters, and the universe as a whole),
and formation of the first stellar systems depended on thermonuclear
reactions between (strongly interacting) protons in the baryon fraction
of the matter density in the universe. This suggests the mass of the
smallest gravitationally bound systems that become stellar systems
at redshift $z$ can be estimated by setting the escape velocity of
protons on the holographic screen for the minimum mass stellar system,
with radius $R_{min}$, equal to the average velocity of protons in
equilibrium with CMB radiation outside the screen. For $R>R_{min}$,
the escape velocity (escaping proton temperature) on the holographic
screen is such that escaping protons are at higher temperature than
the CMB and can transfer heat (and energy) to the CMB. Correspondingly,
for $R<R_{min},$the escape velocity (escaping proton temperature)
on the holographic screen is such that escaping protons are at lower
temperature than the CMB and cannot transfer heat (and energy) to
the CMB. Any protons outside the holographic screen for the minimum
mass stellar system that are in equilibrium with the CMB (such as
those escaping from structures larger than minimum size) can transfer
heat (and energy) to structures less than minimum size until they
grow to minimum size.

The escape velocity for a proton of mass $m_{p}$ gravitationally
bound at radius $R$ from the centroid of a structure with mass $M$
is calculated from $\frac{1}{2}m_{p}v^{2}=\frac{GMm_{p}}{R}$. If
the escape velocity of a proton on the holographic screen for the
minimum mass stellar system at redshift $z$ is the velocity of a
proton in thermal equilibrium with the CMB, $\frac{3}{2}kT=\frac{GMm_{p}}{R}$,
where the CMB temperature $T=(1+z)2.725^{o}K$ and the Boltzmann constant
$k=1.38\times10^{-16}$(g cm$^{2}$/sec$^{2}$)/$^{o}K$. Since the
radius $R$ of the holographic screen for a structure of a mass $M$
is $R=\sqrt{\frac{M}{0.16}}$, the minimum mass of a stellar system
at redshift $z$ is $M_{stellar}=\frac{1}{0.16}\left(\frac{1.5k(1+z)2.725}{Gm_{p}}\right)^{2}$.
If outgoing protons near the holographic screen are in thermal equilibrium
with the CMB and the outgoing photon flow from the minimum mass star,
the outgoing photon flow from stellar systems with mass less than
the minimum stellar system mass is at lower temperature than the CMB
and cannot transfer energy to the CMB or appear as a star against
the CMB background. Note that radii of holographic screens for stellar
systems are considerably larger than radii of stars themselves. For
example, the radius of the holographic screen for our sun is comparable
to the radius of the entire solar system including the Oort cloud.

If the number of structures $n\left(m\right)$ in a mass bin $m$
is $n\left(m\right)=\frac{K}{m}$, the smallest scale structures are
most numerous. The mass of the largest known star is about $6.4\times10^{35}$g
\cite{key-11}. This holographic analysis suggests stellar systems
with mass $6.4\times10^{35}$g would be the minimum mass stellar structures
and the most numerous luminous structures in the universe at $z\approx62$,
consistent with indications that the first stars formed at $z\approx65$
\cite{key-12}. Today, at $z=0$, this analysis indicates the smallest
stellar systems have 0.08 times the solar mass, consistent with the
mass of the smallest stars \cite{key-13}. The fact that the mass
of the smallest stars can be estimated from the extended holographic
principle using only the Boltzmann constant, CMB temperature, gravitational
constant and proton mass suggests a relation between organization
of information and gravity, electromagnetism and strong interactions
underlying that embodied in specific equations modeling details of
thermonuclear reactions and stellar dynamics. 

When matter dominates, the speed of pressure waves affecting matter
density at redshift $z$ within structural level $i$ is $c_{si}(z)=c\sqrt{\frac{4(1+z)^{4}\rho_{r}(0)}{9\rho_{i}(z)}}$
\cite{key-7}, and the Jeans' length at that level $L_{i+1}(z)=c_{si}(z)\sqrt{\frac{\pi}{G(1+z)^{3}\rho_{i}(z)}}$
\cite{key-7}. The first level of large scale structure within the
universe is determined by the Jeans' mass $M_{1}(z)=\frac{4\pi}{3}\left(\frac{L_{1}(z)}{4}\right)^{3}\rho_{0}(z)$,
where $L_{1}(z)=\frac{\left(1+z\right)^{2}}{\rho_{0}(z)}\frac{2c}{3}\sqrt{\frac{\pi\rho_{r}(0)}{G}}=\frac{(1+z)^{2}B}{\rho_{0}(z)}$,
and $B=\frac{2c}{3}\sqrt{\frac{\pi\rho_{r}(0)}{G}}=2.89\times10^{-3}\frac{g}{cm^{2}}$,
so the resulting Jeans' mass $M_{1}(z)=M_{1}=\frac{\pi B^{3}}{48\rho_{0}^{2}(0)}$
is independent of $z$ \cite{key-7}. Evolution of large scale structure
is characterized by $N(z)$ , the number of structural levels between
the Jeans' mass $M_{1}$ and stellar systems, and $n(z)$, the average
number of next lower level structures within a structure at any given
level, as structures in the $N(z)$ levels coalesce into the three
levels present today. The Jeans' mass $M_{i}(z)$ of structures in
level $i$ is determined by the Jean's length $L_{i}(z)$ in the next
highest structural level and the holographic density $\rho_{i-1}(z)$
inside the holographic screen for the Jeans' mass $M_{i-1}(z)$ of
the next highest structural level. So, the ratio of the Jeans' mass
$M_{i}(z)$ to the Jeans' mass $M_{i+1}(z)$ in the next subordinate
level is $\frac{M_{i}(z)}{M_{i+1}(z)}=\frac{L_{i-1}^{3}(z)\rho_{i-1}(z)}{L_{i}^{3}(z)\rho_{i}(z)}=\frac{\rho_{i}^{2}(z)}{\rho_{i-1}^{2}(z)}$.
The holographic density $\rho_{i}(z)=\frac{3A}{4\pi R_{i}(z)},$where
$A=0.16\frac{g}{cm^{2}}$ and the radius of the holographic screen
for the Jeans' mass $M_{i}(z)$ is $R_{i}(z)=\sqrt{\frac{\pi B^{3}(1+z)^{6}}{48A\rho_{i}^{2}(z)}}.$
So, $\frac{M_{i}(z)}{M_{i+1}(z)}=\frac{\rho_{i}^{2}(z)}{\rho_{i-1}^{2}(z)}=\left(\frac{3A}{\pi B}\right)^{3}\frac{1}{(1+z)^{6}}=\frac{1.37\times10^{5}}{(1+z)^{6}}$.
The average mass $\overline{M_{i}(z)}$ of structures in level $i$
is the total mass of the next lowest level of structures within level
$i$ divided by the total number of next lowest level of structures
within level $i$. So, $\overline{M_{i}(z)}=\left(\int_{M_{i+1}(z)}^{M_{i}(z)}m\frac{K}{m}dm\right)/\left(\int_{M_{i+1}(z)}^{M_{i}(z)}\frac{K}{m}dm\right)=M_{i}(z)\left(1-\frac{M_{i+1}(z)}{M_{i}(z)}\right)/\left(\ln\left(\frac{M_{i}(z)}{M_{i+1}(z)}\right)\right)$.
Then, the number $n(z)$ of average mass structures of next lower
level within the average mass at any structural level is $n(z)=\frac{\overline{M_{i}(z)}}{\overline{M_{i+1}(z)}}=\frac{M_{i}(z)}{M_{i+1}(z)}=\left(\frac{3A}{\pi B}\right)^{3}\frac{1}{\left(1+z\right)^{6}}=\frac{1.37\times10^{5}}{(1+z)^{6}}$.

The growth of $n(z)$ tracks development of self-similar large scale
structure. Self-similar large scale structures began to emerge when
$n\approx10$ at $z=3.9$, with 16 structural levels exceeding the
minimum stellar system mass of 2$M_{\bigodot}$. As time went on,
$n=100$ at $z=2.3$ with eight structural levels exceeding the minimum
stellar system mass of 0.9$M_{\bigodot}$, $n=1000$ at $z=1.3$ with
five structural levels exceeding the minimum stellar system mass of
0.4$M_{\bigodot}$, and $n=10,000$ at $z=0.55$ with four structural
levels exceeding the minimum stellar system mass of 0.2$M_{\bigodot}$.

This analysis allows quick simulation of the formation of self-similar
large scale structures, since the number $N(z)$ of self-similar structural
levels exceeding the minimum stellar system mass $M_{min\, stellar}(z)$
is the integer truncation of $\frac{1}{log(\frac{M_{i}}{M_{i+1}})}log(\frac{M_{1}}{M_{min\, stellar}(z)})$,
and the number of average mass structures of next lower level within
the average mass at any structural level, is $n(z)=\left(\frac{3A}{\pi B}\right)^{3}\frac{1}{\left(1+z\right)^{6}}=\frac{1.37\times10^{5}}{(1+z)^{6}}$.

Some other comparisons with observations are worth noting. First,
combining the virial theorem with the holographic relation $M=0.16R^{2},$
the average root mean square velocity of subelements in a self-similar
large scale structure of mass $M$ within the universe is $v_{rms}=\sqrt{\frac{G}{2}}\left(0.16M\right)^{\frac{1}{4}}$.
For an average supercluster mass of $2.2\times10^{49}$g, the r.m.s
galaxy velocity is $2.5\times10^{8}$cm/sec. This compares favorably
with the estimated $4.8\times10^{8}$cm/sec closing velocity of the
colliding {}``bullet cluster'' galaxies 1E0657-56 \cite{key-14}.
Second, the extended holographic principle can be used to derive a
relation between angular momentum of large scale structures and their
mass, similar to that found by Wesson \cite{key-15}. The angular
momentum $J=I\omega$, where the moment of inertia $I$ of a spherical
system of mass $M$ is $I=\frac{2}{5}MR^{2}$, and $\omega$ is the
angular velocity of the system. Using the holographic relation $M=0.16R^{2}$
yields $J=\left(\frac{2}{5}\right)0.16M^{2}\omega$. The angular velocity
can be determined by considering a mass $m$ fixed on the surface
of the rotating structure just inside the holographic screen for the
structure, with radius $R_{s}$. The radial acceleration of that particle
$a_{r}=-\omega^{2}R_{s}$ results from the gravitational force $F_{r}=-\frac{GmM}{R_{s}^{2}}$
attracting the particle to the centroid of the structure, so $\omega^{2}=\frac{GM}{R_{s}^{2}}=\frac{G}{\sqrt{0.16M}}$.
The result is $J=p(M)M^{2}=\frac{2}{5}\frac{G^{0.5}}{(0.16M)^{0.25}}M^{2}$.
Then, $p(M)=1.5\times10^{-15}$ for an average galactic mass of $1.5\times10^{44}$g,
about twice Wesson's empirical value $p=8\times10^{-16}$ \cite{key-15}.

Finally, Forbes and Kroupa \cite{key-16} suggest galaxies and star
clusters can be distinguished by their relaxation times, with galaxies
having relaxation times greater than the age of the universe and star
clusters having relaxation times less than the age of the universe.
Based on standard texts (Shu \cite{key-17} and Binney \& Tremaine
\cite{key-18}), Bhattacharya \cite{key-19} considers a system of
mass $M$ and radius $R$ composed of $N$ stars with average mass
$m$ and number density $n=\frac{3N}{4\pi R^{3}}$. He then approximates
the two body relaxation time for the system as $t_{R}\approx\frac{0.1N}{\ln N\sqrt{Gmn}}$.
Using the holographic relation $R=\sqrt{\frac{M}{0.16}}$ between
the mass and the radius of a system, its relaxation time is $t_{R}\approx\frac{0.1}{\ln N}\sqrt{\frac{4\pi N}{3Gm}}\left(\frac{M}{0.16}\right)^{\frac{3}{4}}$.
This extended holographic analysis indicates the average star cluster
today has mass $1.2\times10^{39}$g. If the (imprecisely known) mass
of the average star is the solar mass $2\times10^{33}$g, the relaxation
time for an average mass star cluster is $8.54\times10^{17}$sec.
If the age of the universe is $13.6\times10^{9}$yr $=4.29\times10^{17}$sec
and the average stellar mass is about twice the solar mass, the relaxation
time of the average mass star cluster equals the age of the universe.
This indicates star clusters have relaxation times of the order of
the age of the universe or less, and larger mass structures have longer
relaxation times. So, a direct consequence of the extended holographic
principle and the fact that the average stellar mass is near the solar
mass is that relaxation times for galaxies are greater than the age
of the universe, consistent with Forbes and Kroupa \cite{key-16}.
\end{doublespace}

\begin{doublespace}

\section{Conclusion}
\end{doublespace}

\begin{doublespace}
The above analyses, based on four simple assumptions, produce numerical
results in general agreement with astrophysical observations of large
scale structures in our universe. It is unlikely that all of these
results are mere coincidence, so the four assumptions probably provide
a reasonable basis for studying development of large scale astrophysical
structures if our universe turns out to be closed.
\end{doublespace}

\end{document}